\documentclass[preprint]{aastex}
\newcommand{\co}{$^{13}$CO}
\newcommand{\kms}{km~s$^{-1}$}
\newcommand{\msun}{M$_{\sun}$}
\newcommand{\SST}{$Spitzer~Space~Telescope$}

\begin{document}
\title{The Discovery of a Massive Cluster of Red Supergiants with
GLIMPSE} \author{Michael J. Alexander\altaffilmark{1}, Henry
A. Kobulnicky\altaffilmark{1}, Dan P. Clemens\altaffilmark{2},
Katherine Jameson\altaffilmark{2}, April Pinnick\altaffilmark{2}, and
Michael Pavel\altaffilmark{2}}

\altaffiltext{1}{Department of Physics and Astronomy, University of
Wyoming, Laramie, WY 82071} \altaffiltext{2}{Institute for
Astrophysical Research, 725 Commonwealth Ave., Boston University,
Boston, MA 02215}

\slugcomment{Draft of 2009 March 06}

\begin{abstract}
We report the discovery of a previously unknown massive Galactic star
cluster at $\ell=29.22$\degr, $b=-0.20$\degr. Identified visually in
mid-IR images from the {\it Spitzer} GLIMPSE survey, the
cluster contains at least 8 late-type supergiants, based on followup
near-IR spectroscopy, and an additional 3--6 candidate supergiant
members having IR photometry consistent with a similar distance and
reddening. The cluster lies at a local minimum
in the \co\ column density and 8 $\mu$m emission. We interpret this 
feature as a hole carved by the energetic winds of the evolving massive
stars.  The \co\ hole seen in
molecular maps at $V_{LSR}\sim95$ \kms\ corresponds to near/far
kinematic distances of 6.1/8.7$\pm$1 kpc.
We calculate a mean spectrophotometric distance of
$7.0^{+3.7}_{-2.4}$ kpc, broadly consistent with the kinematic distances
inferred.  This location places it near the
northern end of the Galactic bar.  
For the mean extinction of $A_V=12.6\pm0.5$ mag ($A_K=1.5\pm0.1$
mag), the color-magnitude diagram of probable cluster members is well
fit by isochrones in the age range 18--24 Myr. The estimated cluster 
mass is $\sim20,000$~M$_\odot$. With the most massive original cluster
stars likely deceased, no strong radio emission is detected in this
vicinity.  As such, this RSG cluster is representative of adolescent
massive Galactic clusters that lie hidden behind many magnitudes of
dust obscuration.  This cluster joins two similar red supergiant
clusters as residents of the volatile region where the end of our
Galaxy's bar joins the base of the Scutum-Crux spiral arm, suggesting
a recent episode of widespread massive star formation there.

\end{abstract}


\section{Introduction}

The \SST\ and the Galactic Legacy Infrared Mid-Plane Survey
Extraordinaire (GLIMPSE) \citep{be03} have opened new windows into
many aspects of star formation and Galactic structure. The GLIMPSE
mid-IR survey and many complementary surveys such as MIPSGAL
\citep{ca05} at 24 $\mu$m, 2MASS \citep{sk06}, and the Boston
University Five College Radio Astronomy Observatory (BU-FCRAO)
Galactic Ring Survey (GRS) in \co\ \citep{ja06} together constitute
powerful new probes at 1--30 arcsecond resolutions of previously
obscured components of the Milky Way.  These overlapping surveys have
produced many new discoveries including a new globular cluster
\citep{ko05} and new young open clusters \citep{me05,fi06,da07}. A
more complete census of star clusters enables a better understanding
of the Galaxy's formation and structure, its star formation history,
and its current rate of star formation.

Massive stars are rare and short-lived. A typical 9 \msun\ star will
spend about 26 Myr on the main-sequence, while a 15 \msun\ star has a
MS lifetime of around 13 Myr \citep{me00}. After leaving
the main-sequence these stars begin their helium burning red
supergiant (RSG) phase. The He-burning lifetime is about 4 Myr for 9
\msun, 2.5 Myr for 15 \msun\ and even shorter for higher mass stars
\citep{me00}. This is a very small window for observing these evolved
stars and the reason why there are relatively few of them known in the
Galaxy. \citet{be94} discovered 5 RSGs in NGC~7419, and more recently 
\citet{fi06} discovered 14 RSGs and a yellow (G-type) supergiant in 
single Galactic cluster. The current record holder for the most number
of RSGs in a single Galactic cluster is Stephenson~2, also known as RSGC2 
\citep{st90,da07}, which has 26 cluster RSGs. The number of massive 
RSGs in a cluster places strong limits on cluster age (massive stars 
have not all disappeared yet), cluster mass (enough initial mass to produce
large numbers of high mass stars), and the duration of the star
formation burst that produced the cluster, in that the stars must have
formed in a relatively short period to have reached the same evolved state
simultaneously.  \citet{fi06} and \cite{da07} use an evolved
initial-mass function (IMF) to derive initial cluster masses of 20,000
-- 40,000 M$_\sun$ for RSGC1 and RSGC2.

In this paper, we describe the discovery of another RSG cluster using
methods similar to those of \citet{fi06}. The cluster was discovered
serendipitously as a conspicuous grouping of bright red stars while
perusing 3-band mid-infrared color mosaics from GLIMPSE. At 
$\ell=29.223$\degr\ and $b=-0.207$\degr, the cluster lies in roughly
the same direction as the RSG clusters reported in \citet{fi06} and 
\citet{da07}. We use archive data from $Spitzer$, 2MASS, and the GRS, 
along with new near-IR spectroscopy of nine probable members to 
constrain the stellar content, age, and distance of this young and 
massive Galactic cluster.

\section{Datasets Employed}

\subsection{Infrared Archival Data}

Our discovery and analysis make use of archival data from the \SST\
GLIMPSE~I/II and MIPSGAL~I legacy surveys.  The GLIMPSE \citep{be03}
project was conducted using the Infrared Array Camera (IRAC)
\citep{fa04} wavebands centered near 3.6, 4.5, 5.8, and 8.0
$\micron$.\footnote{GLIMPSE and MIPSGAL data are available from the
$Spitzer$ Science Center website $http://ssc.spitzer.caltech.edu/$}
With four seconds of integration time per target, 5$\sigma$
sensitivities of 0.2, 0.2, 0.4, and 0.4 mJy were achieved for each
band, respectively, on point sources with angular resolutions of
$\sim$2\arcsec. In total, the GLIMPSE~I implementation covers
$\mid\ell\mid$ = 10--65\degr\ and $\mid~b\mid$ $<$1\degr.  Each image
was calibrated by the {\it Spitzer Science Center} and further
processed by the GLIMPSE team pipeline to produce point source
catalogs (PSCs) and 3\degr$\times$2\degr\ mosaics.  The \SST\ MIPSGAL
survey \citep{ca05} covers the same region of sky as GLIMPSE but uses
the Multi-band Infrared Photometer for Spitzer (MIPS) \citep{rieke}
bandpasses at 24, 70, and 160 $\micron$.  The 24 $\micron$ mosaics 
have a resolution of 6\arcsec\ and a 5$\sigma$ sensitivity of 
1.7 mJy \citep{ca05}.

\begin{figure}
\includegraphics[width=18cm]{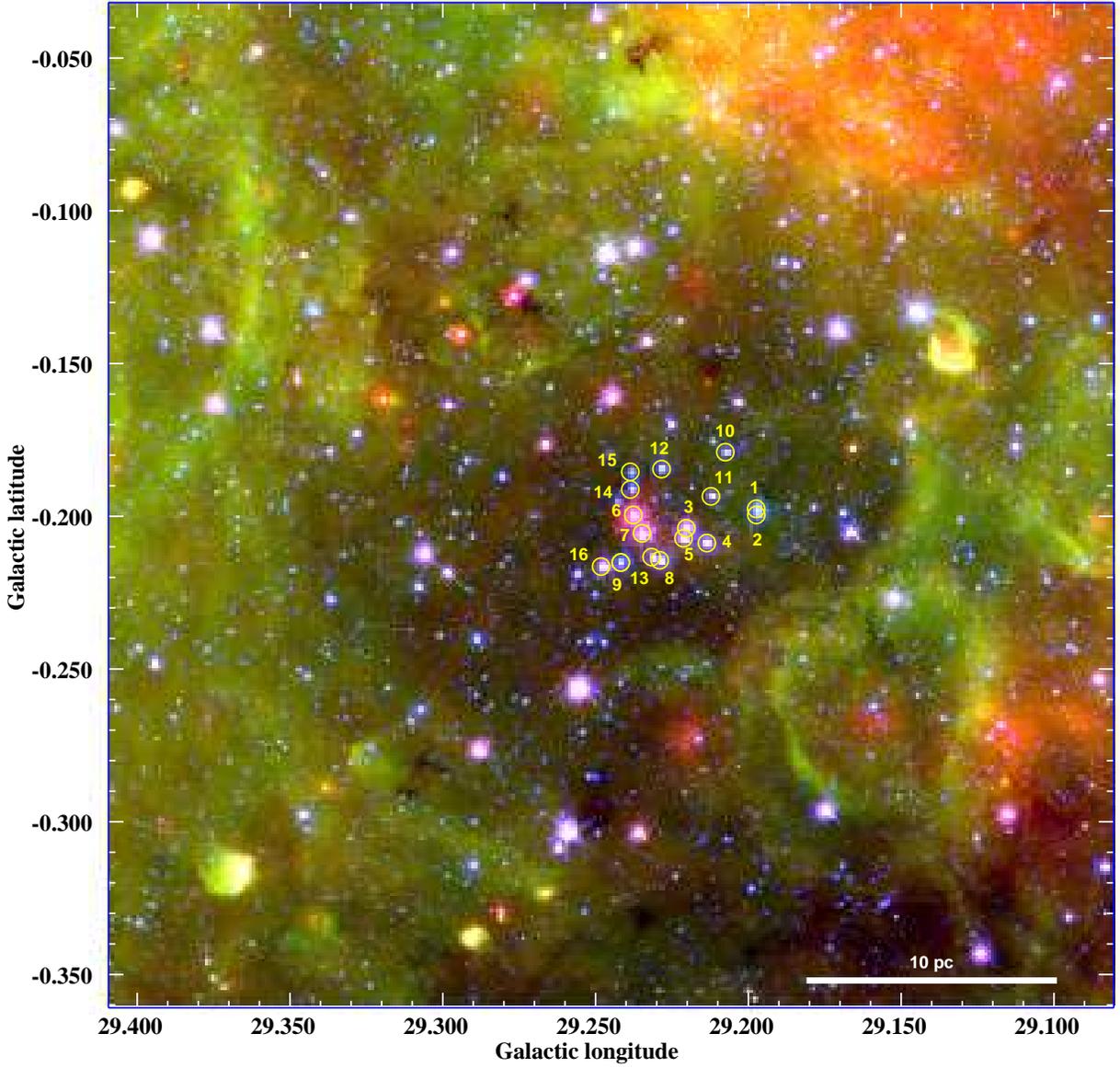}
\caption{Three color IRAC \& MIPS image of the cluster region with 
4.5 $\micron$ in blue, 8.0 $\micron$ in green, and 24 $\micron$ in red.
Numbered circles 1--9 denote cluster stars with followup
near-IR spectroscopy, while 10--16 are candidate cluster0 stars near
the cluster core.  The bar at lower right shows a linear scale 
of 10 pc at the adopted cluster distance of 7.0 kpc. \label{cluster}}
\end{figure}

Figure~\ref{cluster} shows a 3-color image of the star cluster field,
with $Spitzer$ IRAC [4.5] in blue, IRAC [8.0] in green, and MIPS [24]
in red. Blue predominantly highlights the stellar photospheres.  Green
shows diffuse emission from warm molecular cloud interfaces, i.e., PAH
emission from photodissociation regions excited by UV photons.  Red
preferentially reveals warm dust. The scale bar in the lower right shows
a linear scale of 10 pc at a distance of 7.0 kpc.  The grouping of
bright stars near the center of the image suggests a physical 
association.  They also lie near a local  minimum in the diffuse 
emission, possibly a cavity sculpted by the winds and 
energetic photons from young massive stars. Infrared dark clouds 
(IRDCs) and dark filaments appear superimposed on the bright diffuse 
background at many locations across this field, indicating the 
presence of opaque, cold clouds in the foreground. A handful of very 
red and bright sources, often coinciding with dark clouds, surround 
the cluster.  The color and placement of these objects suggest that 
they are young stellar objects (YSOs) associated with the IR 
emission.

We also use photometry from the University of Massachusetts Two Micron
All Sky Survey (2MASS) \citep{sk06}, which observed the sky
in the near infrared for J (1.25 $\micron$), H (1.65 $\micron$),
and K$_s$ (2.16 $\micron$) bands. The project attained magnitude
limits of J = 15.9, H = 15.0, and K$_s$ = 14.3, with a signal-to-noise
ratio of 10.

\begin{figure}
\includegraphics[width=18cm]{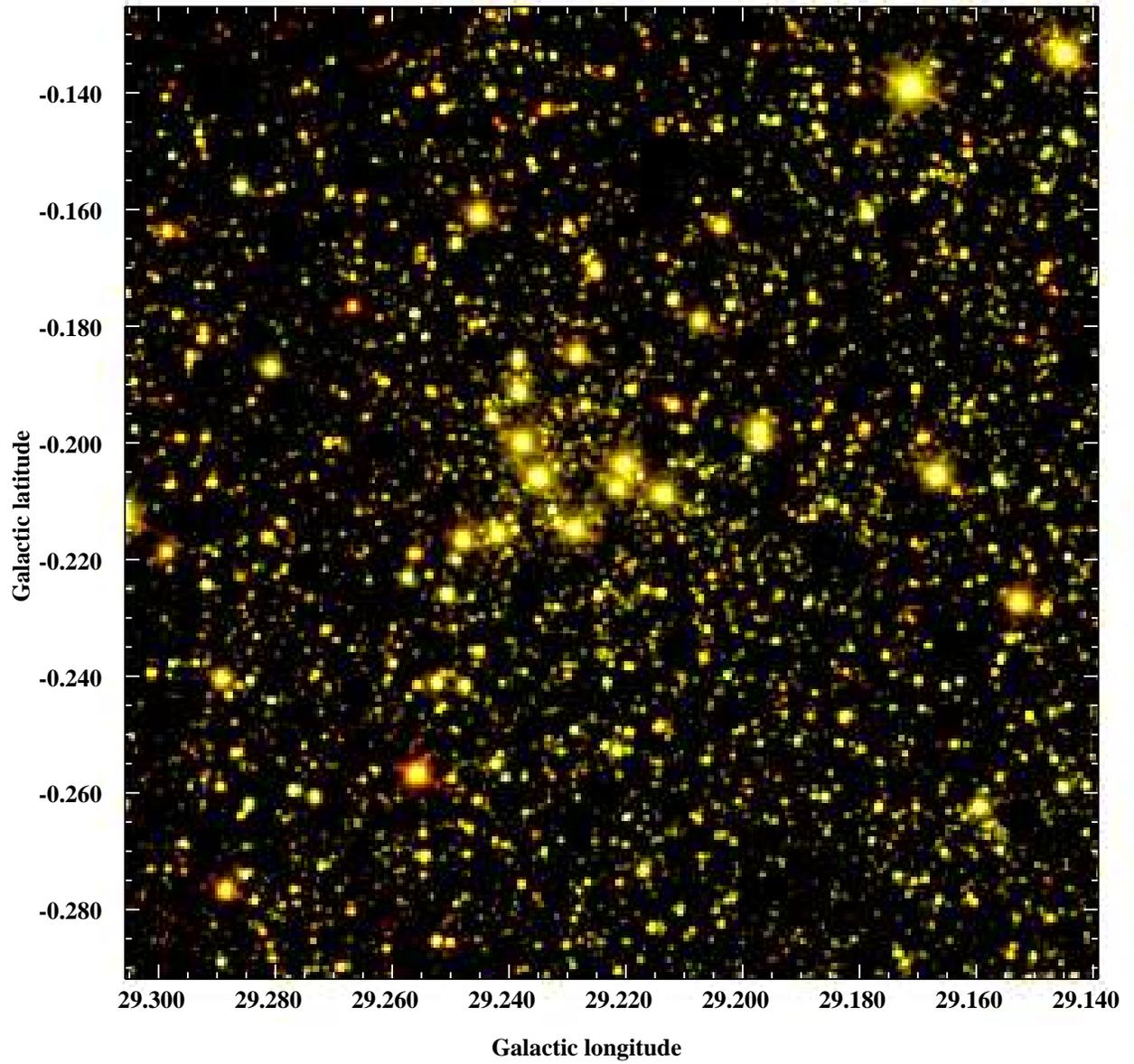}
\caption{An enlarged three-color image of the cluster field with
2MASS J band in blue, H band in green, and $K_s$ band in red.  \label{2mass}}
\end{figure}

Figure~\ref{2mass} shows a color composite image of the cluster with
the 2MASS J band in blue, the H band in green, and the $K_s$ band in
red. The brightest cluster stars are conspicuous as a grouping of 
yellow-red stars near the center of the image.

Nine stars (numbered 1 through 9 in Figure~\ref{cluster}) were 
chosen for spectroscopic observation based on their location 
(clustering) and brightness in the GLIMPSE/MIPSGAL mosaics. 
Table~\ref{phot} lists these nine stars along with their equatorial 
coordinates, 2MASS JHK and GLIMPSE IRAC photometry. Most stars are 
saturated in the IRAC bands.

\subsection{Near-IR Spectroscopy}

Near-infrared spectroscopy was conducted using the Mimir instrument
\citep{cl07} on the Perkins 1.8~m telescope, located outside
Flagstaff, Arizona, on the UT nights of 2008 May 21, June 25--28, and
September 23. Mimir is a cryogenic reimager with spectroscopic
capability, delivering up to a 10$\times$10 arcmin field of view to a
1024$\times$1024 InSb ALADDIN III array detector at 0.6 arcsec per
pixel. Read noise is typically 17--18 e$^-$ RMS.

All spectroscopy used the JHK-grism (120 $l$~mm$^{-1}$, 30\degr\ blaze
angle, CaF2 substrate with 29\degr\ wedge) in conjunction with one of
three long-pass (LP) filters and the wide-field, F5, camera optics to
yield spectra with resolutions of 430--780, for J-band through K-band, 
respectively.  All May and some June observations used the 1.16 $\mu$m
LP filter to yield spectra unaliased to 2.27 $\mu$m, while the 
remaining June observations use the 1.90 $\mu$m LP to form spectra to 
2.5 $\mu$m.  September observations used a 1.4 $\mu$m LP to 
simultaneously observe the full H and K bands.

The slit used has projected sizes of 1.2 arcsec (2 pixels) by 5 arcmin
and was oriented North-South.  Spectra were typically exposed for
15--45 seconds then moved along the spatial direction and another
spectrum obtained. This A-B cycle was repeated for 5 or 6 pairs.  The
dome was rotated to cover the telescope beam, and exposures were taken
with continuum lamps on and off in order to produce spectral
flat-fields. Spectral images of an argon lamp positioned over the slit
were obtained for wavelength calibration. For calibration of the
wavelength dependence of atmospheric transmission, spectroscopy of
bright A0V stars was performed using all of the same steps, for stars
chosen to be within 15 degrees and 0.1 airmasses of the target stars.

Data reduction used custom IDL-based programs that included measuring
and correcting the non-linear response of the InSb photovoltaic
pixels, using the fourth-order method described in \citet{cl07}, in
addition to the usual dark and flat-field corrections. Wavelength
calibration utilized the argon spectra to ascertain the 2-D
spatial-dispersion coefficients necessary to remap the spectral images
into uniformly dispersed versions.  Typical dispersion solution
scatter was at or under 0.1 pixels (0.14 nm) for all argon
spectra. Target spectral images were flattened, interpolated, and
remapped to 20 subpixels for each raw pixel, then rebinned to twice
the original dispersion to achieve a fixed channel spacing and common
reference wavelength.

Spectra were modeled on the images as sloped lines with ``bow tie''
spatially-broadened extrema, due to the grism action.  Spectra were
extracted from the modeled locations by fitting locally sloped
backgrounds at each spectral channel outside the spatial extent covered
by the spectra, then integrating the signal above the background over
the 2--3 rows showing significant emission. Noise was estimated for
each extracted spectral channel based on gaussian read noise and
poisson background and signal contributions. This somewhat
overestimates the noise for these 2:1 synthetic channels. Relative
scaling factors for each of the 10--12 spectra for a target were
found across either the H- or K-band. For each spectral channel, the
10--12 scaled values were tested to remove bad pixels or cosmic rays,
using median filtering. The surviving data were averaged using
weighting by inverse noise squared.

Telluric spectra were developed using the A0V standard star extracted
spectra and the SpeXTool program XTelcor by \citet{cush04} and used to
correct the target star spectra.  The final spectra were flattened by
identifying wavelengths free of lines, fitting baselines, and dividing
by the baseline functions to yield relative intensity spectra.  The
resulting spectra, normalized by their error-propagated noise spectra,
showed signal-to-noise ratios of 200 - 450.

\subsection{\co\ Molecular Line Data}
As a means of assessing the molecular environment of the cluster, we
used data cubes from the BU-FCRAO GRS\footnote{GRS data can be
obtained from the Boston University website
$http://www.bu.edu/galacticring/$.} \citep{ja06}, which mapped the
\co\ J = 1 $\rightarrow$ 0 molecular line in the first Galactic
quadrant from $l$ = 18\degr\ to 55.7\degr\ and $\mid b\mid$ $\le$
1\degr. The FCRAO 14 m telescope has a beamsize of 46\arcsec\ FWHM; 
the spectra were smoothed to a velocity resolution of 0.21 \kms, and 
cover a local standard of rest (LSR) velocity range from -5 to 135 
\kms.  The RMS channel sensitivity of the GRS is 0.4~K $T_A^*$.

\section{Analysis}

\subsection{Near-IR Spectra}

\begin{figure}
\includegraphics[width=18cm]{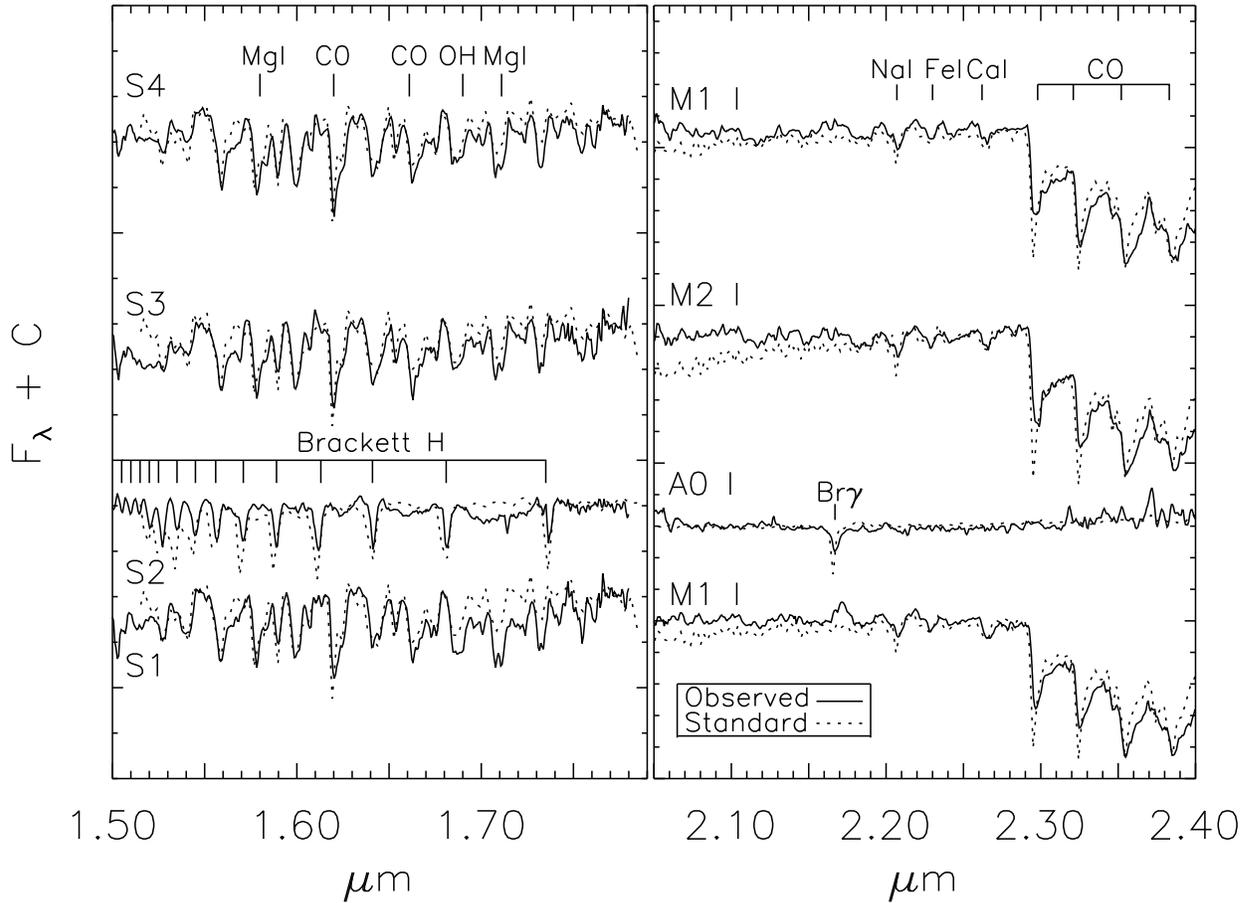}
\caption{Spectra for stars S1 through S4.  The wavelengths cover
approximate H- and K-band ranges. Our observations are the solid black
lines, while smoothed atlas spectra are shown as dotted lines.
The spectral types for the atlas stars are given, and the most
prominent spectral features are labeled.  \label{spec1}}
\end{figure}

\begin{figure}
\includegraphics[width=18cm]{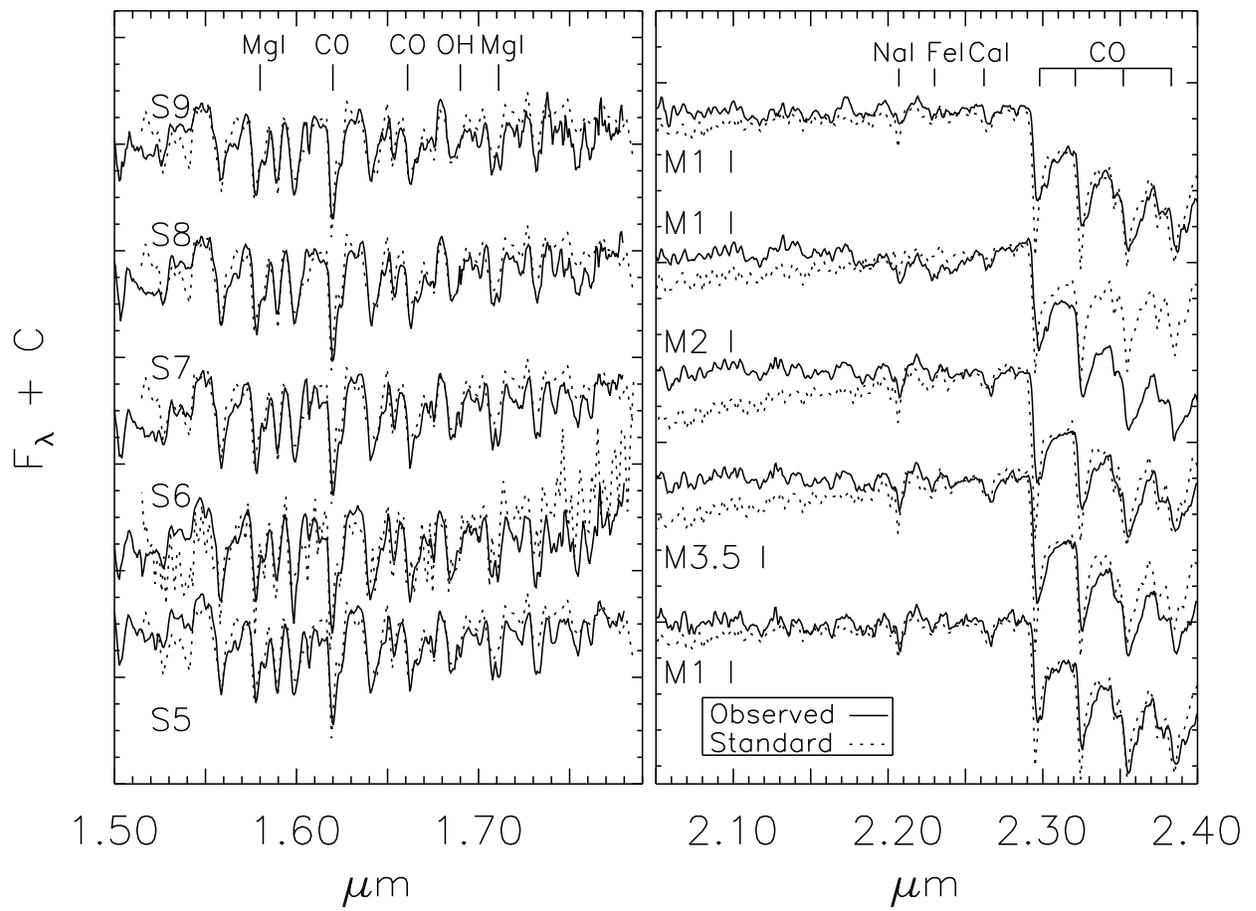}
\caption{Spectra for stars S5 through S9, as in Figure~\ref{spec1}.
\label{spec2}}
\end{figure}

Near-infrared spectra of the first nine potential cluster members are 
displayed in Figures~\ref{spec1}~\&~\ref{spec2}.  Solid lines show 
portions of the normalized H-band and K-band spectra for each star, 
shifted by arbitrary offsets for clarity. Dashed lines show H-band 
spectra from \citet{me98} and K-band spectra from \citet{wa97}
\footnote{The data were obtained from the NOAO Digital Library 
$http://www.lsstmail.org/dpp/library.html$.} for the nearest available
spectral type. The spectra of \citet{me98} and \citet{wa97} have a 
resolution R=3000, which we smoothed by a factor of 10 to match the 
resolutions of our data. Eight of the nine stars (S1 and S3 through S9)
show deep CO bandhead features near 2.3 $\mu$m, characteristic of cool,
evolved stars.  These spectra are most consistent with late-type
giants or supergiants.

Star S2, unlike the rest, is best matched with an early A or late B
supergiant spectrum. This determination is based on the presence of 
strong Brackett series features in the H-band and Brackett $\gamma$ 
near 2.17 $\mu$m. The H-band Brackett lines of S2 were only about
half as wide as the same lines in the A0V comparison stars, confirming
its low-gravity, supergiant status. The lack of molecular
absorption features (e.g., CO) is another indication that S2 is an
early-type star.

\begin{figure}
\includegraphics[width=18cm]{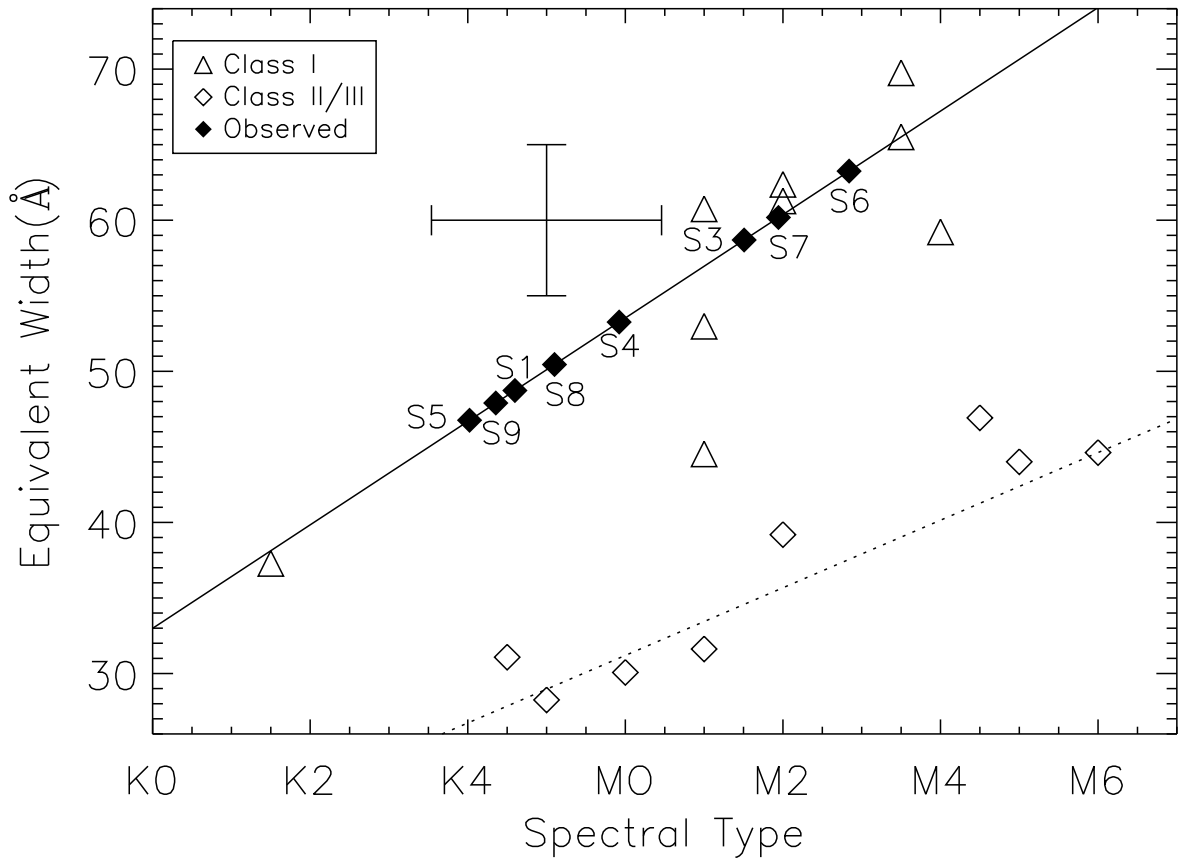}
\caption{Equivalent width of the CO bandhead versus spectral type.
Open triangles show supergiants and open diamonds show giants
from the stellar spectral atlases of \citet{me98} and \citet{wa97}.
The solid and dashed lines are least squares fits to the
supergiant and giant sequence, respectively.
Filled diamonds are our sources, placed along the best-fitting line
for supergiants.  The cross represents our uncertainties of $\pm$ 
5~\AA~ and $\pm$ 2 sub-types for each star.
\label{EW}}
\end{figure}

Following the method outlined by \citet{fi06} (see their Figure~8), we
measured the equivalent width (EW) of the first $^{12}$CO overtone
feature at 2.30 $\mu$m for each star (except S2 as it shows no CO
absorption). EW values were found by setting the continuum level to be
the average value of the normalized flux between 2.28 and 2.30
$\micron$ and integrating the absorption between 2.30 and the
pseudocontinuum longward of 2.32 $\micron$.  The same method was used
for the observed stars as well as stars with known spectral types
from \citet{wa97}. Figure~\ref{EW} shows the correlation between EW
and spectral type for the atlas stars of supergiant ({\it open
triangles}) and giant ({\it open diamonds}) luminosity classes.  The
solid line is a least-squares fit to 14 G3--M6 supergiants, and the dashed
line is a least-squares fit to 8 G7--M4 giants. This linear dependence
extends all the way down to G3 spectral classes, however, we
only show spectral types  K0 to M7 in Figure~\ref{EW} for clarity. 
The giants show systematically lower EW than supergiants for a given 
spectral type. The strong correlation between EW and spectral type, 
along with the systematic differences between giants and supergiants, 
shows that even at relatively low spectral resolution, our IR spectra
can be used to discern spectral and luminosity classes. Typical values
for the EW of cluster members are 45 to 65 \AA, which we used to 
assign the spectral types according to the best-fit line; these are 
shown as solid diamonds in Figure~\ref{EW}. We have estimated our 
uncertainty on the EW to be $\pm$~5 \AA\, and given the slope of the 
least-squares fit, this corresponds to $\pm$~2 sub-types. 
Table~\ref{phot} lists these adopted spectral types.

\subsection{IR Photometry}

\begin{figure}
\includegraphics[width=18cm]{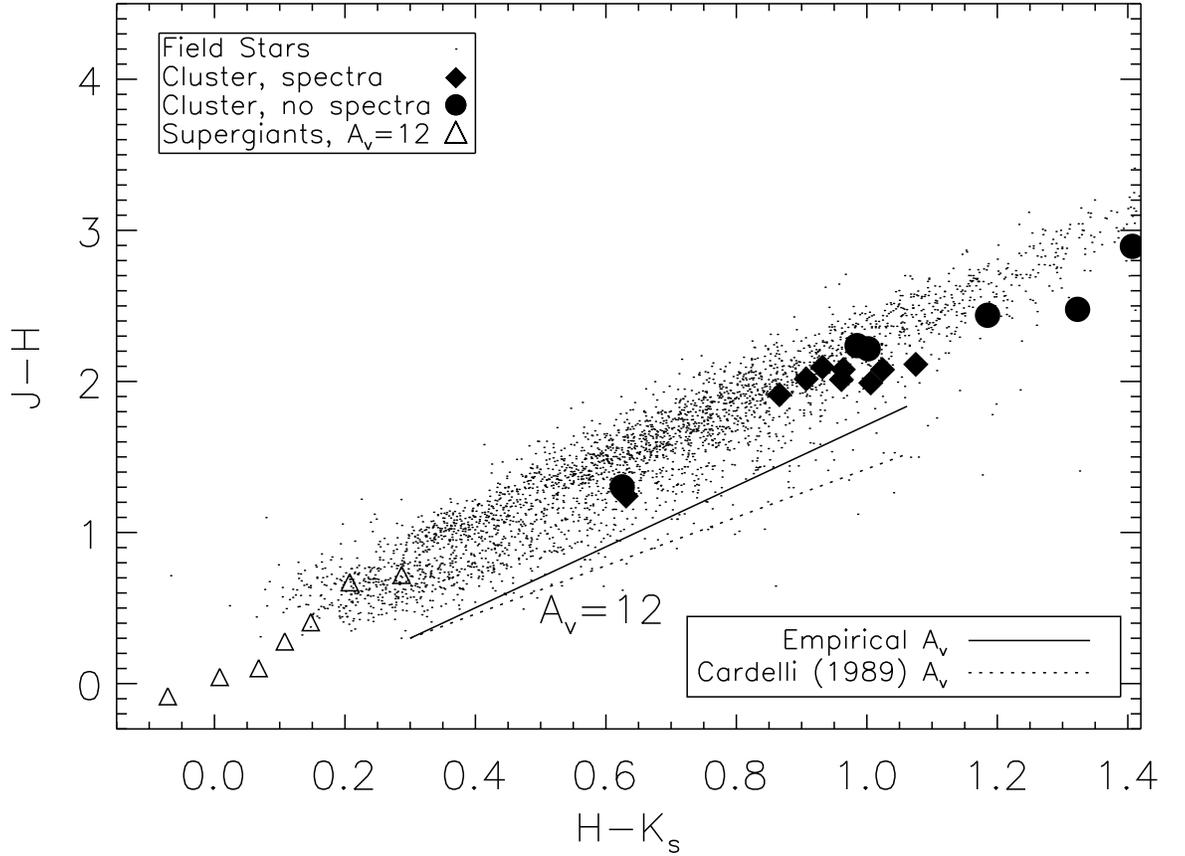}
\caption{2MASS J-H vs. H-K$_s$ color-color diagram of the field in
Figure~\ref{cluster}.  Small dots designate every point source in
Figure~\ref{cluster} with accurate ($\sigma<0.3$ mag) JHK$_s$ photometry.
Filled diamonds designate the nine stars with near-IR spectra.  
Filled circles denote other stars near the cluster core with K$_s$ < 8.  
Open triangles mark the fiducial unextincted supergiant sequence. 
The solid line shows a reddening vector for $A_V$ = 12~mag.  
\label{ccdms}}
\end{figure}

Figure~\ref{ccdms} is a 2MASS J-H vs. H-K$_s$ color-color diagram
showing the nine stars with spectral types ({\it filled diamonds}), 
bright (K$_s<8$) stars within $\sim$4 arcmin of the cluster center 
({\it filled circles}), and field stars in Figure~\ref{cluster} 
({\it dots}). Open triangles with labels denote standard JHK$_s$ 
colors for supergiant stars.\footnote{The photometric data is in the 
2MASS JHK$_s$ color system, so we used transformations from 
\citet{ca01} and \citet{be88} to convert the data from \citet{wh80}, 
\citet{el85} and \citet{ko83} to the 2MASS filter system. } The dotted
 line is the reddening vector using the \citet{ca89} extinction law,
and the solid line is an empirical reddening vector based on the slope
of the locus of points. This was done to attempt to correct for the 
deviation in the data from the \citet{ca89} reddening. The difference 
between the slopes corresponds to $\Delta$(J-H) = 0.25 over 12 magnitudes 
of visual extinction. Since J-H offset is small compared to our 
estimated error, we still adopt the \citet{ca89} reddening law for our
data analysis.  Eight of the nine stars with near-IR spectral 
measurements occupy a limited region of the color-color diagram --- 
$<$J-H$>$ $\simeq$ 2.0, $<$H-K$_s>$ $\simeq$ 1.0 --- consistent with similarly
reddened late-type supergiants. S2 has a much different color than the 
rest of the stars, suggesting an earlier spectral type. Of the other 
bright stars without spectral types, S14 is similar to S2 and less red
than most cluster members, while S11, S13, and S16 are more red, 
suggesting that they may be more heavily extincted. Stars S10 and S12 
have similar colors to the majority of supergiants, suggesting that 
they may also be RSG cluster members. Using the reddening law of 
\citet{ca89}, we converted the 2MASS colors, J--K$_s$ and H--K$_s$,
into an average extinction. We find a mean extinction of $A_V=12.6\pm0.5$ 
(A$_{K_s}=1.5\pm0.1$).

\begin{figure}
\includegraphics[width=18cm]{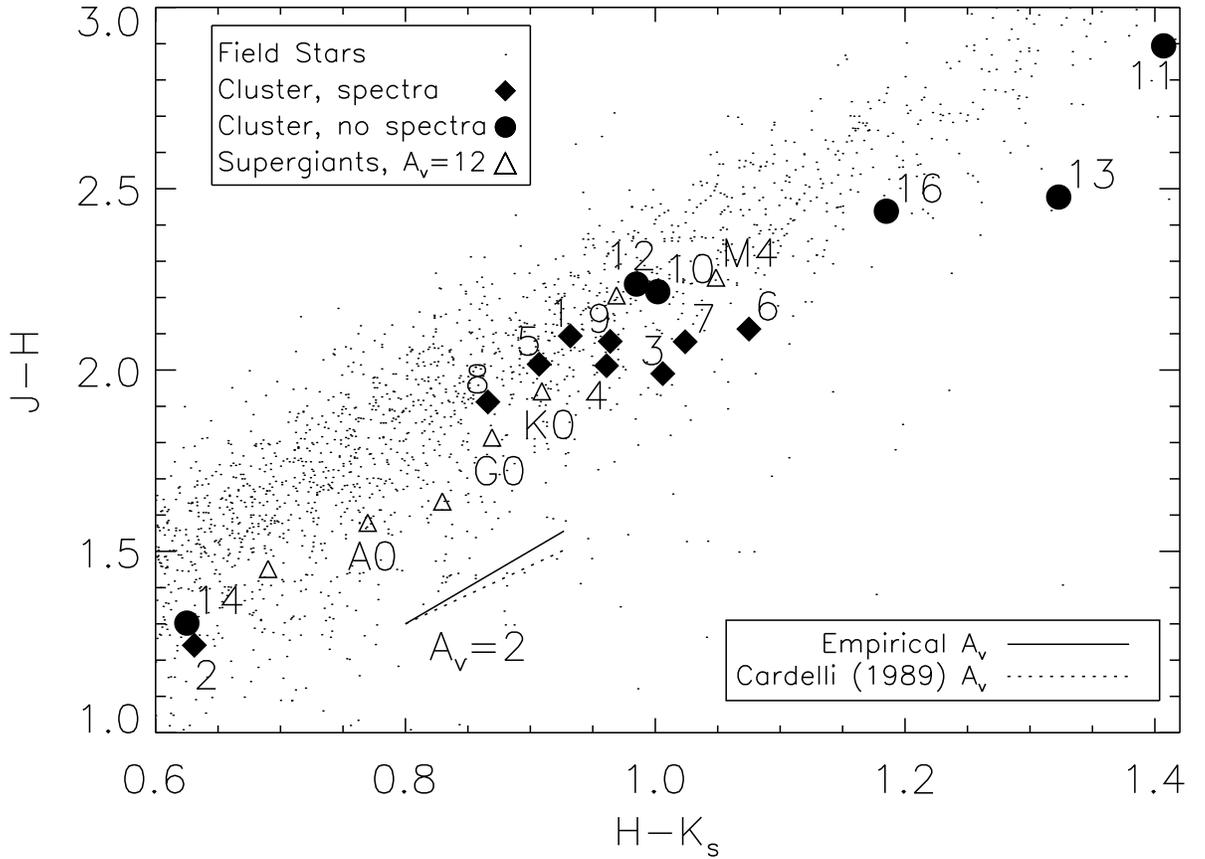}
\caption{Enlarged 2MASS color-color diagram as in Figure~\ref{ccdms}, 
with the fiducial supergiant sequence reddened by $A_V$ = 12. 
A reddening vector for $A_V$ = 2 is shown. The reddened M-type supergiant
standards match the colors of the cluster stars. S2, an A0I,
is less red than expected and is probably a foreground object.  \label{ccd}}
\end{figure}

Figure~\ref{ccd} is an enlarged version of Figure~\ref{ccdms}, with
the fiducial supergiant colors reddened by $A_V=12$ using the 
empirical relation metioned previously.  Most of the
stars show colors matching K through M supergiants, with some 
differential extinction among cluster members. However, S2 does not 
match the expected A0I supergiant at H--K~$_s$~=~0.77 and 
J--H~$_s$~=~2.1. S2 appears to be less reddened by 1--2 magnitudes. 
This indicates that S2 (and S14) are likely foreground objects.

\subsection{Molecular Environment and a Kinematic Distance}

We used the GRS \co\ data cubes to survey the molecular environment
near this RSG cluster. Figure~\ref{vlsr} shows a plot of the \co\
brightness temperature versus LSR velocity averaged across the field
of view pictured in Figure~\ref{cluster}.  This figure shows two
strong \co\ peaks at LSR velocities of roughly 79 and 95 \kms\, and
several smaller peaks at 8, 47, 64, and 101 \kms.  The abundance of
\co\ components shows that this is a complex sightline with many
molecular complexes at various distances.  Numbers above each peak
indicate the corresponding near and far heliocentric kinematic
distances of $d_{kin}$=5.0/9.8 kpc and 6.1/8.7 kpc for the two major
peaks and 0.7/14.1 kpc, 3.2/11.6~kpc, 4.2/10.6~kpc, and
$\sim$7.4~kpc\footnote{This is the unique tangent-point distance
corresponding to 103 \kms.} for the lesser peaks, respectively. The 
numbers in parentheses above each peak indicate the Galactocentric 
distance for the given velocity. These distances are derived from the 
Galactic rotation curve of \citet{cl85}, assuming $R_0=8.5$ kpc, 
$\theta_0=$220 \kms.  The more recent rotation curve of \citet{le08} 
yields very similar distances for the same $R_0$ and $\theta_0$.  
Adopting $R_0=7.6$~kpc instead \citep{eisenhauer} would reduce all 
distances by about 10\%. We esitimate the kinematic distance uncertainty
at $\pm$ 15\% based on the width of the \co\ features.

Maps of integrated intensity were created
for each of the five spectral features and compared to the
GLIMPSE images. Four of the features showed no correlation
between the CO and GLIMPSE images. For the fifth feature, with a peak
at 95 \kms, a strong correlation of the CO distribution with the 
GLIMPSE 8 $\mu$m  distribution was seen, as was a distinct 
anticorrelation with the location of the stellar cluster. 

In Figure~\ref{hole}, we overlay a map of the GRS \co\ brightness 
temperature at 95 \kms, averaged between 86--101 \kms, on the GLIMPSE images.
The GLIMPSE [4.5] and [8.0] bands are shown in blue and green,
respectively, while the \co\ is shown in red, along with white
contours.  This velocity range exhibits a striking absence of
molecular gas at the location of the cluster while the surrounding
regions are \co-bright. This \co\ minimum coincides with the 8.0
$\micron$ hole discussed earlier in connection with
Figure~\ref{cluster}.  We interpret this feature to be a hole in the
ISM formed by the winds of massive stars that have already evolved
over the lifetime of the cluster. This velocity range, which peaks at
94 \kms, has a near kinematic distance of 6.1 kpc and a far kinematic 
distance of 8.7 kpc.  Distinguishing between these two
possiblities requires an independent distance measurement. 

\subsection{A Spectrophotometric Cluster Distance}

Spectroscopic parallax is commonly used to find distances when the 
relation between spectral type and absolute magnitude is well-known 
and has small dispersion (e.g., for main sequence stars).  Red 
supergiants, however, exhibit a large range of luminosities, spanning
several magnitudes at any given spectral type.  This is seen in the 
evolutionary tracks from \citet{ma08}, which are nearly vertical at 
the coolest temperatures. The large dispersion of abolute magnitudes
among red supergiants in \citet{le05} supports this empircally. In 
order to estimate a spectrophotometric distance for the cluster, 
we calculated the mean absolute V magnitude and standard deviation for
RSG spectral types K3 through M4.5 (representative of our sample) 
from \citet{le05}. By adopting a mean intrinsic V-K color of 4.2 
\citep{le05}, we derive a mean absolute K magnitude for red supergiants of
$M_K=-10.0~\pm~0.9$. Given our mean observed K magnitude of 5.75 and
extinction $A_K=1.5$ for our sample (excluding the possible foreground
star S2), we estimate a distance of $7.0^{+3.7}_{-2.4}$ kpc. 


Since the spectrophotometric distance is highly uncertain, we can only 
broadly constrain the distance of the cluster using spectroscopic
parallax to the broad range 4 -- 11 kpc, which encompasses both 
kinematic distances. For the purposes of determining a cluster age 
below, we will adopt the spectrophotometric distance of 7.0 kpc.

\subsection{Cluster Age}

\begin{figure}
\includegraphics[width=18cm]{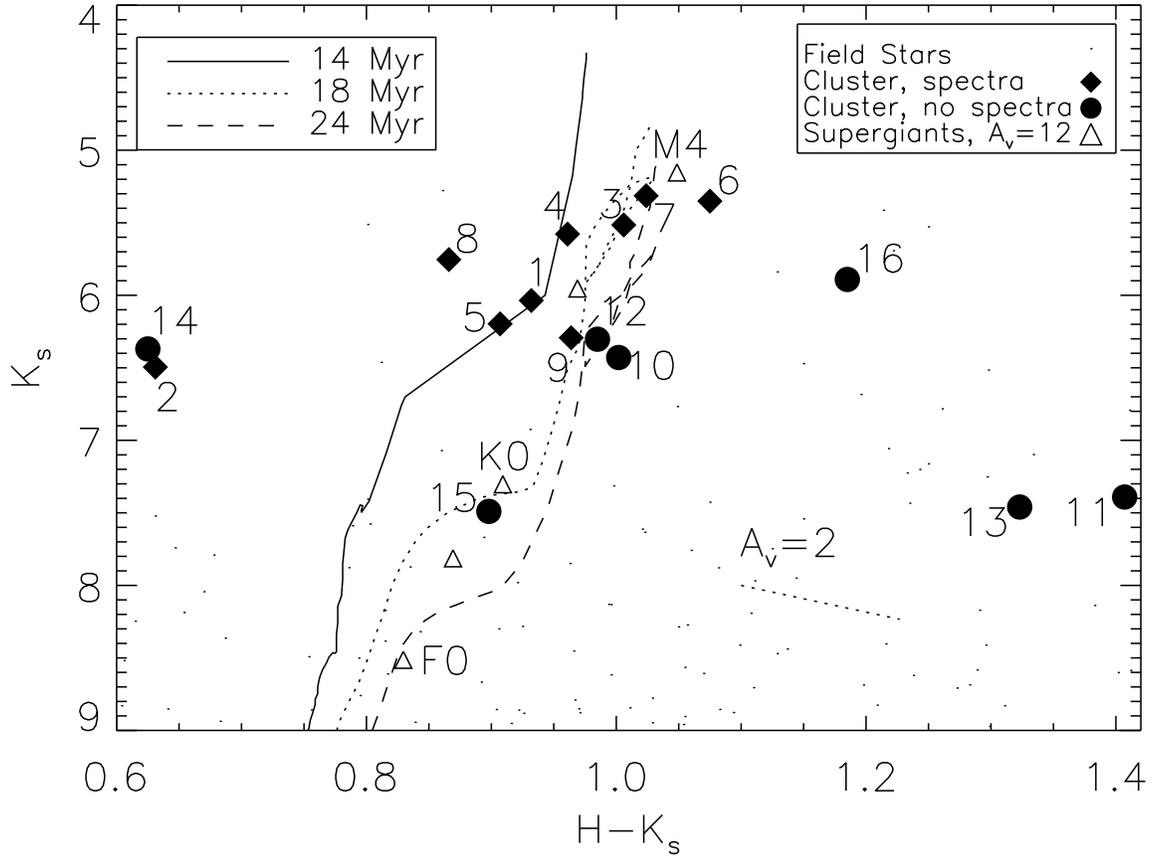}
\caption{Color-magnitude diagram of stars in the field of
Figure~\ref{cluster}, with labeling as in Figure~\ref{ccdms}.  Reddened
\citep{ca89} isochrones with ages of 14 Myr ({\it solid}), 18 Myr ({\it
dotted}), and 24 Myr ({\it dashed}) at a distance of 7.0 kpc are
shown. M-type supergiants match many of the cluster
stars, while S2 is less red and more than 2.5 magnitudes brighter
than a comparable A supergiant star (off figure) at the same distance.
The best fit age is somewhere between 18 and 24 Myr.  Note
the presence of several stars (S10, S12, S 15) with photometry but lacking spectroscopy
({\it filled circles}) that have similar colors and magnitudes as
confirmed M supergiants, suggesting the presence of several more
supergiant members. S14 was shifted up by 0.1 mag for clarity\label{cmd}}
\end{figure}

\begin{figure}
\includegraphics[width=18cm]{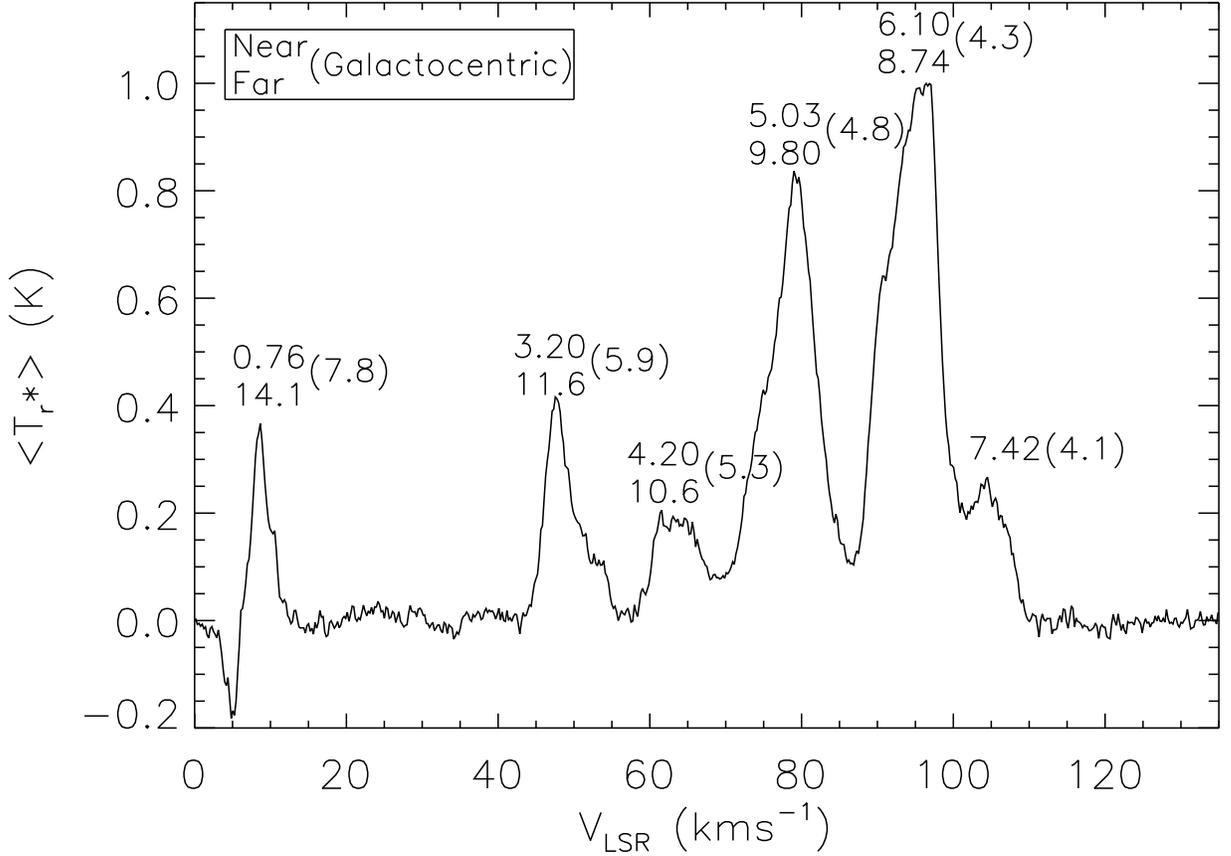}
\caption{\co\ spectrum averaged over the area shown in
Figure~\ref{cluster}, showing at least five \co\ peaks along
this sightline.   Numbers above each peak denote the near and far
heliocentric kinematic distances in kpc based on the \citet{cl85} rotation
curve, while numbers in parentheses indicate the corresponding
Galactocentric distance.  We identify the cluster with the molecular
feature near 95 \kms\ (see text).
 \label{vlsr}}
\end{figure}

\begin{figure}
\includegraphics[width=18cm]{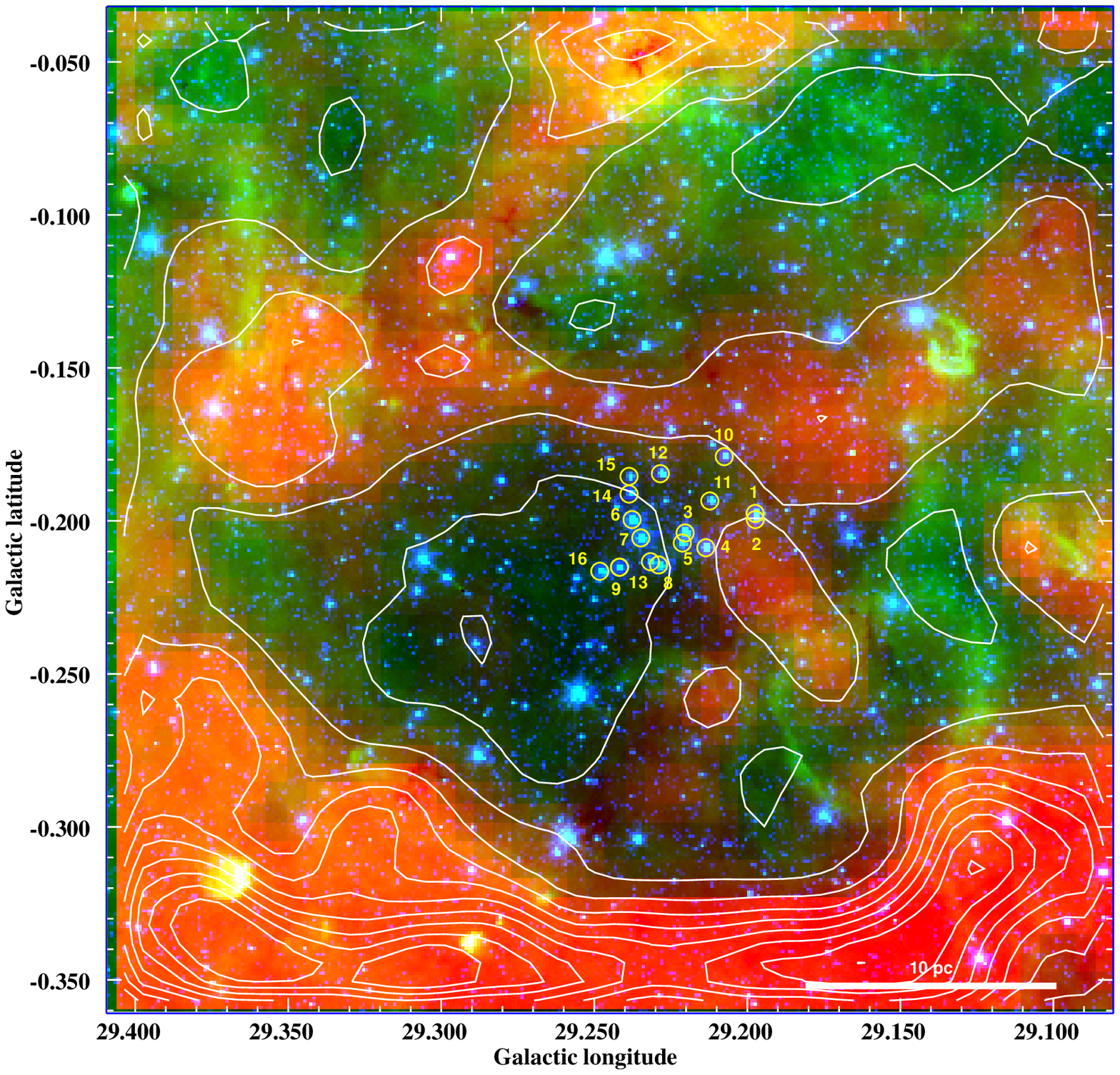}
\caption{Three color image of the cluster region
with 4.5 $\micron$ in blue, 8.0 $\micron$ in green, and the GRS \co\ data
averaged over the velocity range  86--101~\kms\
($d_{kin}\simeq$6.1~kpc/8.7~kpc) in red with \co\ contours in white.
Contours range from $T_A^*$=0.1~K to $T_A^*$=20.1~K in increments of
2.0~K. The cluster lies near the \co\ minimum, suggestive of
 a cavity blown by the stellar winds
of the massive cluster members. \label{hole}}
\end{figure}

Figure~\ref{cmd} shows a K$_s$ vs. H-K$_s$ color-magnitude diagram, 
along with the fiducial supergiant sequence, at a distance of 7.0 kpc 
and reddened by A$_V$ = 12. Symbols are the same as Figure~\ref{ccdms}.
The solid, dotted, and dashed curves are 14, 18, and 24 Myr 
isochrones\footnote{Isochorones obtained from 
$http://stev.oapd.inaf.it/cmd$} from \citet{ma08}. They have been 
reddened by A$_V$ = 12 and placed at a distance of 7.0 kpc. The majority 
of the cluster stars are consistent with being reddened supergiants 
at 7.0 kpc for ages between 14 and 24 Myr.
Note that two objects with photometry but lacking spectroscopy, S10 and
S12, fall in the same region as spectroscopically confirmed RSGs, 
bringing the total number of probable red supergiants in the cluster 
to ten. Another object, S15, closely matches the G0I fiducial point.
There are five other candidates, S2, S11, S13, S14, and S16 that lie 
very close to the center of the cluster spatially, but do not occupy the same
color-magnitude space. The A0 I star, S2, is much less red and nearly
two magnitudes brighter than it should be if it were a cluster member. 
As stated previously we consider S2 to be a foreground star, not associated
with the cluster. Based on the same criteria (but lacking spectroscopy)
S14 is also likely to be a foreground object. Stars S11, S13, and S16 
may be background objects or possibly cluster members enshrouded 
by additional intracluster or circumstellar dust. \citet{massey05} 
show that supergiants may be surrounded by up to 5 visual magnitudes 
of circumstellar extinction. Thus, we consider stars S11, S13, and S16 
as candidate cluster supergiants, given their brightness and proximity 
to the cluster core. From Figure~\ref{cmd} we are then able to conclude 
that the cluster contains 8 spectroscopically confirmed supergiants as well as 6 
additional cluster candidates.

The most massive stars remaining in each isochrone are 18.6 \msun, 14.9
\msun, and 12.6 \msun~ corresponding to ages of 14 Myr, 18 Myr,
and 24 Myr. These stars are the most luminous objects in the cluster
in the near-IR, so the excellent agreement between the tips isochrones
and the cluster supergiants indicates that the data are consistent 
with an age range of 18--24 Myr. Examination of the the 20-cm radio 
continuum map of this region from the NRAO VLA Sky Survey 
\citep{condon} shows no sources of emission at the location of the 
cluster, consistent with an older age and lack of ionizing photons 
from the most massive stars.

\subsection{Cluster Stellar Mass}

The eight confirmed and 3--6 additional candidate red supergiants in 
the cluster imply a large total stellar mass, and, given the short 
relative lifetime of the RSG phase, a short star formation timescale.
\citet{da08} compute Monte-Carlo population synthesis models with a 
Salpeter IMF \citep{sa55} and Geneva isochrones \citep{me00} to 
estimate the number of red supergiants as a function of age for a range
of initial cluster masses. Based on their Figure~8, a cluster 
containing 10 RSGs at an age of 18--24 Myr has a initial mass of 
$\sim$20,000 M$_\odot$.  In this cluster, all stars except the most 
luminous supergiants are faint due to its distance and reddening, 
which means the underlying cluster is below the detection threshold 
for 2MASS.

\section{Discussion and Conclusions}

This newly discovered cluster resembles two other RSG clusters in 
terms of stellar mass, age, the high number of red supergiants, and 
placement in the Galaxy.  Table~\ref{summary} summarizes the
locations and derived parameters for the clusters RSGC1
\citep{fi06,da08}, RSGC2~$\equiv$~Stephenson~2 \citep{st90,da07}, and
the present cluster.  We have listed the distances
for each cluster based on published data and computed the
Galactocentric distance, $R_{Gal}$, based on a solar Galactocentric
distance of 7.6 kpc \citep{eisenhauer}.  Although the uncertainties of
the spectrophotometric distances are several kiloparsecs, the resulting
locations at $\ell=25-29$~\degr and $R_{gal}=$ 3.5--3.7 kpc indicate
that these clusters may lie within a few hundred parsecs of each
other.  \citet{da07} suggest that this is where the Scutum-Crux spiral
arm meets the Galactic bulge and is within a stellar ring proposed by
\cite{bertelli}.  This location may also lie near the northern end of
the Galactic bar, the parameters of which are still uncertain. There 
is evidence for at least two non-axisymmetric structures in the 
inner Galaxy, a vertically thick triaxial bulge/bar of half-length 
3.5~kpc and angle 20--35 degrees \citep{bissantz02, gerhard01, babu} 
and a vertically thinner ``Long Bar'' of half-length 4.4 kpc and angle
40--45 degrees \citep{hammersley00, be05}.  RSGC1, RSGC2, and this new
cluster lie near the end of the Long Bar, perhaps where it meets the
Scutum-Crux spiral arm and or the Molecular Ring.

\begin{figure}
\includegraphics[width=18cm]{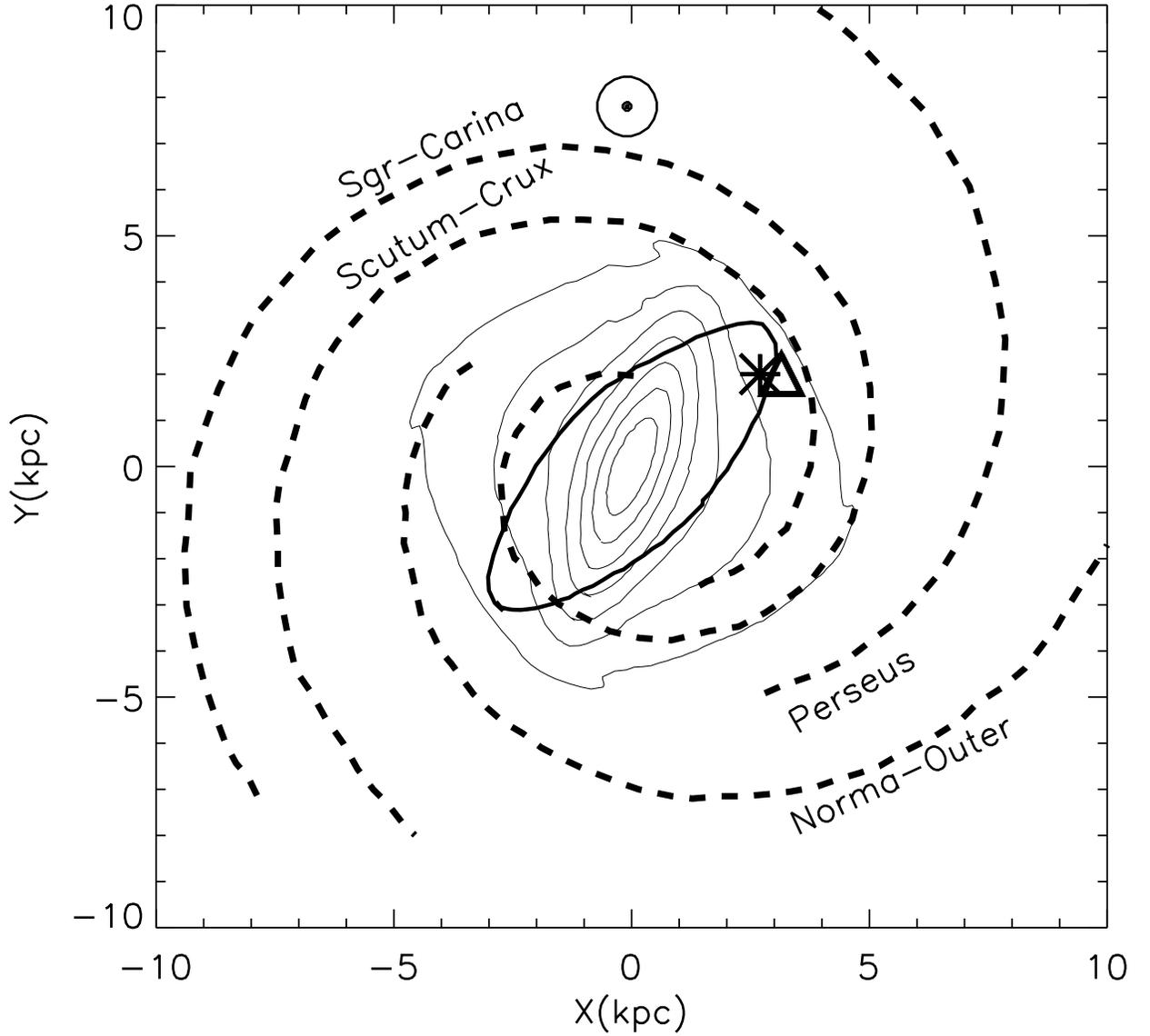}
\caption{Schematic of face-on Milky Way showing the locations of RSGC1
and RSGC2 ({\it asterisk}) and the newly discovered cluster 
({\it triangle}). The symbol sizes represent the uncertainties in the 
clusters' locations. Thick dashed curves denote spiral arms as described 
by  \citet{nakanishi}. Thin contours show the triaxial/bulge bar as 
modeled by \citet{bissantz02}, while the thick solid curve shows the
thin ``Long Bar'' as derived by \cite{be05}. \label{galaxy}}
\end{figure}

Figure~\ref{galaxy} plots the locations of RSGC1 and RSGC2 ({\it
asterisk}) and the newly discovered cluster ({\it triangle}). Thick
dashed lines show the spiral arms as defined by \citet{nakanishi}. 
Thin contours show the stellar bar as modeled by \citet{bissantz02} 
from 2MASS near-IR photometry, while the thick solid curve shows the 
4.4 kpc radius bar inferred by \cite{be05} from \SST\ mid-IR 
photometry. The location of all three RSG clusters near the end of the
bar is consistent with recent star formation, possibly induced by 
radial gas flows along the bar.  Numerical simulations show 
radial flows resulting from barred potentials \citep{schwarz81,
combes85,engl,piner,rod08}.  Observational studies of star formation
in other galaxies also reveal enhanced activity associated with bars,
sometimes along the bars themselves and sometimes near the ends of 
bars where spiral arms or ``spurs'' attach to the bars 
\citep{mf97,sh02}. In cases where star formation occurs along bars, it
is displaced toward the leading side of the stellar bar and peaks 
outside of the dust and molecular lanes that run along the bar 
\citep{mf97,sh02}. For example, \citet{do96} present CO maps of the 
barred spiral galaxy NGC~1530 showing concentrations of molecular 
material near the bar ends at $\sim$4 kpc from the nucleus. The 
accompanying optical image appears to show enhanced star formation 
just outside this region near the bar ends.  \citet{sh00} report that 
H$\alpha$ emission peaks near the ends of the $\sim$4 kpc stellar bar 
in NGC~5383.  The locations of the RSG clusters in Figure~\ref{galaxy} 
near one end of the Milky Way's bar at a $\sim$4~kpc radius hint that 
a similar phenomenon may be happening in our galaxy.

Taken together, the discovery of this new cluster raises the total
count of RSGs in the Galaxy and adds evidence for a localized burst of
recent massive star formation.  The accumulating data point toward 
widespread, high-mass star formation in this vicinity over the last
$\sim$20 Myr.  With a stellar mass of roughly 20,000 M$_\odot$ and a
population containing 8--14 red supergiants, this ranks among the
more massive clusters in the Galaxy.

\acknowledgments We would like to thank Charles Kerton \& Kim
Arviddson for assisting with the initial discovery of this cluster and
for software help.  We acknowledge Bob Benjamin and Danny Dale for
helpful comments.  We thank an anonymous referee for a rapid and
expert review.  HAK and MJA were supported by NASA through grant
NAG5-10770.  This publication makes use of data products from the Two
Micron All Sky Survey, which is a joint project of the University of
Massachusetts and the Infrared Processing and Analysis
Center/California Institute of Technology, funded by the National
Aeronautics and Space Administration and the National Science
Foundation.  This publication makes use of molecular line data from
the Boston University Five College Radio Observatory Galactic Ring
Survey (GRS), funded by the National Science Foundation under grants
AST-9800334, -0098562, -0100793, -0228993, \& -0507657.  This research
was conducted in part using the Mimir instrument, jointly developed at
Boston University and Lowell Observatory and supported by NASA, NSF,
and the W.M. Keck Foundation.  This work was partially supported by
the NSF under grant AST-0607500 to Boston University.

{\it Note added in manuscript.} As this paper went to press we became 
aware of the simultaneous discovery of this red supergiant cluster reported
in \citet{cl09}. Their analysis and conclusions closely match our own.

\clearpage
\begin{deluxetable}{ccccrcccccccc}
\tablecaption{Cluster Members and Candidates \label{phot}}
\tablewidth{18 cm}
\tabletypesize{\scriptsize}
\rotate
\tablehead{\colhead{Star \#}& Status\tablenotemark{a} & \colhead{GLIMPSE ID} &  \colhead{R.A.(J2000)} & 
\colhead{Dec.(J2000)} & \colhead{J}& \colhead{H}& \colhead{K$_s$}
& \colhead{[3.6]\tablenotemark{b}}& \colhead{[4.5]\tablenotemark{b}}
& \colhead{[5.8]}&  \colhead{[8.0]}&  \colhead{S.T.}}
\startdata
1  & M & G029.1974$-$00.1975 &  18:45:19.38& $-$03:24:48.1& 9.06&   6.97& 6.04& \nodata& \nodata& 5.24& 5.20& K5 I\\
2  & F & G029.1973$-$00.1995 &  18:45:19.81& $-$03:24:51.6& 8.37&   7.13& 6.50& \nodata& \nodata& 5.70& 5.75& A0 I\\
3  & M & G029.2206$-$00.2037 &  18:45:23.26& $-$03:23:44.0& 8.51&   6.52& 5.51& \nodata& \nodata& 4.45& 4.20& M1 I\\
4  & M & G029.2139$-$00.2087 &  18:45:23.59& $-$03:24:13.8& 8.55&   6.54& 5.58& \nodata& \nodata& 4.73& 4.50& M0 I\\
5  & M & G029.2215$-$00.2075 &  18:45:24.17& $-$03:23:47.2& 9.12&   7.10& 6.20& \nodata& \nodata& 5.22& 4.84& K4 I\\
6  & M & G029.2380$-$00.1999 &  18:45:24.34& $-$03:22:42.0& 8.54&   6.43& 5.35& \nodata& \nodata& 4.19& 4.14& M3 I\\
7  & M & G029.2351$-$00.2059 &  18:45:25.31& $-$03:23:01.0& 8.42&   6.34& 5.31& \nodata& \nodata& 4.08& 4.04& M2 I\\
8  & M & G029.2290$-$00.2147 &  18:45:26.52& $-$03:23:35.2& 8.53&   6.62& 5.75& \nodata& \nodata& 4.92& 4.63& K5 I\\
9  & M & G029.2421$-$00.2154 &  18:45:28.12& $-$03:22:54.5& 9.34&   7.26& 6.29& \nodata& \nodata& 5.49& 5.44& K4 I\\
10 & C & G029.2076$-$00.1790 &  18:45:16.55& $-$03:23:44.6& 9.65&   7.43& 6.43& \nodata& \nodata& 5.57& 5.52&  \nodata\\
11 & C & G029.2121$-$00.1935 &  18:45:20.13& $-$03:23:54.1&11.69&   8.80& 7.39& \nodata& \nodata& 5.71& 5.54&  \nodata\\
12 & C & G029.2286$-$00.1846 &  18:45:20.03& $-$03:22:46.8& 9.53&   7.29& 6.30& \nodata& \nodata& 5.46& 5.43&  \nodata\\
13 & C & G029.2317$-$00.2137 &  18:45:26.60& $-$03:23:24.7&11.26&   8.78& 7.46& \nodata& \nodata& 6.01& 5.81&  \nodata\\
14 & F & G029.2385$-$00.1911 &  18:45:22.53& $-$03:22:26.4& 8.40&   7.09& 6.47& \nodata& \nodata& 6.12& 6.07&  \nodata\\
15 & C & G029.2386$-$00.1854 &  18:45:21.32& $-$03:22:16.2&10.16&   8.39& 7.49& 7.06   & \nodata& 6.66& 6.69&  \nodata\\
16 & C & G029.2480$-$00.2166 &  18:45:29.00& $-$03:22:37.5& 9.51&   7.08& 5.89& \nodata& \nodata& 4.79& 4.61&  \nodata\\
\enddata
\tablenotetext{a}{Membership status.  M denotes a member, F a foreground object, and C a cluster candidate.}
\tablenotetext{b}{Ellipses indicate no listing in the GLIMPSE Point Source Catalog owing to saturation.}
\end{deluxetable}

\clearpage
\begin{deluxetable}{lccccccccccc}
\tablecaption{Known Red Supergiant Clusters \label{summary}}
\tablewidth{21 cm}
\rotate
\tablehead{\colhead{Name} &\colhead{$\ell$}  &\colhead{b}       &\colhead{R.A.(J2000)} &\colhead{Dec.(J2000)} & \colhead{N RSGs}  &\colhead{Age}  &\colhead{D}     &\colhead{$A_V$ } &\colhead{Mass}       &\colhead{$R_{Gal}$}  \\
         \colhead{}       &\colhead{(\degr)} &\colhead{(\degr)} &  \colhead{}        &     \colhead{}      &     \colhead{}    &\colhead{(Myr)}& \colhead{(kpc)}&\colhead{(mag)}  &\colhead{($M_\odot$)} &\colhead{(kpc)}       }
\startdata
this cluster              & 29.2            & $-$0.20        & 18:45:20           & $-$03:24:43           & 8--14           & 18--24        & 7.0         & $\sim$12                & $\sim$20,000    & 3.9           \\
RSGC1\tablenotemark{a}                     & 25.2            & $-$0.15        & 18:37:58           & $-$06:52:53           & 14                & 12$\pm$2      & 5.8,6.5\tablenotemark{c}         & 24                & 30,000 $-$ 40,000 & 3.5             \\  
RSGC2\tablenotemark{b}                     & 26.2            & $-$0.06        & 18:39:20           & $-$06:01:41           & 26                & 17$\pm$3      & 5.8         & 13                & 40,000          & 3.5              \\
\enddata
\tablenotetext{a}{\citet{fi06}}
\tablenotetext{b}{\citet{da07}}
\tablenotetext{c}{\citet{na06}}
\end{deluxetable}

\clearpage

\clearpage

\clearpage

\clearpage

\clearpage

\clearpage
 
\clearpage

\clearpage

\clearpage

\clearpage

\clearpage

\end{document}